# Low Autocorrelation Binary Sequences: Number Theory-Based Analysis for Minimum Energy Level, Barker Codes

Abhisek Ukil

ABB Corporate Research, Segelhofstrasse 1K, Baden 5 Daettwil, CH-5405, Switzerland

Phone: +41 58 586 7034

Fax: +41 58 586 7358

E-mail: abhisek.ukil@ch.abb.com

*Abstract*

Low autocorrelation binary sequences (LABS) are very important for communication applications. And it is a notoriously difficult computational problem to find binary sequences with low aperiodic autocorrelations. The problem can also be stated in terms of finding binary sequences with minimum energy levels or maximum merit factor defined by M.J.E. Golay, $F = \dfrac{N^2}{2E}$, $N$ and $E$ being the sequence length and energy respectively. Conjectured asymptotic value of $F$ is 12.32 for very long sequences. In this paper, a theorem has been proved to show that there are finite number of possible energy levels, spaced at an equal interval of 4, for the binary sequence of a particular length. Two more theorems are proved to derive the theoretical minimum energy level of a binary sequence of even and odd length of $N$ to be $\dfrac{N}{2}$ and $\dfrac{N-1}{2}$ respectively, making the merit factor equal to $N$ and $\dfrac{N^2}{N-1}$ respectively. The derived theoretical minimum energy level successfully explains the case of $N = 13$, for which the merit factor ($F = 14.083$) is higher than the conjectured value. Sequence of lengths 4, 5, 7,

11, 13 are also found to be following the theoretical minimum energy level. These sequences are exactly the Barker sequences which are widely used in direct-sequence spread spectrum and pulse compression radar systems because of their low autocorrelation properties. Further analysis shows physical reasoning in support of the conjecture that Barker sequences exists only when $N \leq 13$ (this has been proven for all odd $N$).

## 1. Introduction

Low (off-peak) autocorrelation binary sequences (LABS) [1–3] have various applications in many communication engineering problems [4–5]. This includes synchronization in communication systems, where it is very much convenient to detect a sequence having a single high-peak and much lower off-peak levels as per the Wiener-Khinchin theorem [18]. Usage of LABS as modulation pulses in radar and sonar ranging has considerable advantages owing to the facts that the target detection capability increases with energy and good range resolution requires signal with a sharply peaked autocorrelation function [4].

Finding such LABS is a notoriously difficult computational problem. The complexity of the optimization problem in LABS is outlined in [4,6,7]. LABS has a long history of computation. Following Golay's conjectures [1–3] on the merit factor [1] of the LABS, there have been many approaches for finding near-optimal solutions. The LABS problem is shown to be equivalent to the low-energy configurations of a specific spin model with long-range 4-spin interactions by Bernasconi [4], De Groot and Würtz [10]. Rapid increase in computational power has resulted in capabilities for performing a higher dimensional exhaustive searches [3,7]. However, exhaustive search more than a space of $10^{30}$ is rather impractical.

For near-optimal searches for LABS, Schotten and Lüke [11] provides a summary of the search approaches. Jedwab [12] did a survey of the approaches for the best merit factor searches. Militzer et al. [9] used evolutionary algorithm to search for LABS with higher merit factors. Reinholz [13] used parallel genetic algorithm for the LABS problem. Dotu and Hentenryck [14] used tabu search method for the LABS problem. Various meta heuristics like shortest descent local search (SDLS) and tabu search (TS), included in a memetic algorithm, were applied by Gallardo et al [15] for searching the near-optimal solutions. From signal processing point of view, Hayashi [15], Yu and Gong [16] applied Hadamard transform-based approach for two level autocorrelations.

In this paper, a number theory-based analysis of the LABS problem is presented. The remainder of the paper is organized as follows. In section 2, the mathematical problem of LABS is presented. In section 3, the number theory-based analysis of the energy levels and the minimum energy level is presented in details. This is followed by the application results using the number theory-based approach in section 4, concluded in section 5. Some supporting mathematical proof is given in the appendix.

## 2. LABS: low autocorrelation binary sequences

We consider binary sequences of length $N$

$$x = \{x_1, x_2, \ldots, x_N\}, \qquad x_i = \pm 1, \tag{1}$$

and their aperiodic autocorrelations,

$$R_k = \sum_{i=1}^{N-k} x_i x_{i+k}, \qquad k = 0, 1, \ldots, N-1. \tag{2}$$

The energy of the sequence is defined as

$$E = \sum_{k=1}^{N-1} R_k^2. \tag{3}$$

The merit factor $F$ of such sequence, as defined by Golay [1,2] is

$$F = \frac{N^2}{2\sum_{k=1}^{N-1} R_k^2} = \frac{N^2}{2E}. \tag{4}$$

Here, the goal is to achieve sequences with $F$ as large as possible, in other words the energy $E$ as low as possible. That is,

arg max $F$,

or (5)

arg min $E$.

Golay described the conjectured asymptotic value of $F = 12.32$ [1] for very long sequences. Jensen and Hoholdt [8] argued the true asymptotic value of the maximum merit factor to be 6 instead of 12.32. However, many sequences [21] have been found with merit factor more than 6.

## 3. Number theory-based analysis

### 3.1 Energy levels

For simplicity, let's start with a smaller sequence with five elements, $x = \{x_1, x_2, x_3, x_4, x_5\}$. So, the aperiodic autocorrelation as per (2) (except $R_0$ which is not considered in energy (3)), for $x$ becomes

$$\begin{aligned} R_1 &= x_1 x_2 + x_2 x_3 + x_3 x_4 + x_4 x_5, \\ R_2 &= x_1 x_3 + x_2 x_4 + x_3 x_5, \\ R_3 &= x_1 x_4 + x_2 x_5, \\ R_4 &= x_1 x_5. \end{aligned} \quad (6)$$

It is to be noted that $R_k$, $(k = 1,2,3,4)$ has $N - k$ elements, $N = 5$ is the sequence length.

So, the energy would be

$$E = \sum_{k=1}^{4} R_k^2 = (x_1x_2 + x_2x_3 + x_3x_4 + x_4x_5)^2 + \\ (x_1x_3 + x_2x_4 + x_3x_5)^2 + (x_1x_4 + x_2x_5)^2 + (x_1x_5)^2 \quad (7)$$

On rearranging (7), we get

$$E = \begin{Bmatrix} (x_1x_2)^2 + (x_2x_3)^2 + (x_3x_4)^2 + (x_4x_5)^2 + \\ (x_1x_3)^2 + (x_2x_4)^2 + (x_3x_5)^2 + \\ (x_1x_4)^2 + (x_2x_5)^2 + \\ (x_1x_5)^2 \end{Bmatrix} + \\ 2\begin{Bmatrix} (x_1x_2x_2x_3) + (x_1x_2x_3x_4) + (x_1x_2x_4x_5) + (x_2x_3x_3x_4) + (x_2x_3x_4x_5) + (x_3x_4x_4x_5) + \\ (x_1x_3x_2x_4) + (x_1x_3x_3x_5) + (x_2x_4x_3x_5) + \\ (x_1x_4x_2x_5) \end{Bmatrix}. \quad (8)$$

$= X + 2Y \; (say)$

From (8), in general, considering the fact that each $R_k, (k = 1, 2, \ldots, N-1)$ has $N-k$ elements, $X$ would have $\sum_{i=1}^{N-1} i = \dfrac{(N-1)N}{2}$ elements. In this case $X$ would have

$\dfrac{(5-1)5}{2} = 10$ elements. And elements in $X$ are of squares of $\pm 1$, therefore all elements of $X$ would be +1. Hence,

$$X = \frac{N(N-1)}{2}. \quad (9)$$

On the other hand, elements of $Y$ come from the multiplication of the combinations of any two elements at each level of $R_k$, $(k = 1, 2, \ldots, N-2)$. If total number of elements in $Y$ be $m$, then

$$m = \sum_{k=1}^{N-2} \binom{N-k}{2} = \frac{N(N-1)(N-2)}{6}, \tag{10}$$

where, $\binom{n}{k}$ indicates the combination operation of the form 'n_choose_k'.

See appendix for the proof of (10). In this example for $N = 5$,

$m = \binom{4}{2} + \binom{3}{2} + \binom{2}{2} = 6 + 3 + 1 = 10$. The energy thus becomes

$$E = X + 2Y = \frac{N(N-1)}{2} + 2Y. \tag{11}$$

Considering the goal of the problem (see (5)), we would like to achieve minimum $E$. From the right-hand side (RHS) of (11), it would be only possible if $Y$ (having $m$ elements, given by (10)) becomes negative by some combinations of the $m$ elements. Following this, we would like to prove the following theorem.

Theorem 3.1. The number of possible energy levels of the binary sequence of a particular length are fixed and finite, with a difference of 4 among the contiguous levels.

*Proof*: For the RHS of the energy equation in (11), let's consider $Y$ with $m$ elements consists of $n$ number of $-1$ and $p$ number of $+1$.

$$m = n + p. \tag{12}$$

Then,

$$Y = -n + p = m - 2n. \tag{13}$$

Using (10 & 13), the energy equation (11) becomes

$$\begin{aligned} E &= \frac{N(N-1)}{2} + 2m - 4n \\ &= \frac{N(N-1)}{2} + 2 \cdot \frac{N(N-1)(N-2)}{6} - 4n \\ &= \frac{N(N-1)(2N-1)}{6} - 4n \end{aligned} \tag{14}$$

For a sequence with particular length $N$, $m$ is fixed and finite (given by (10)). In the RHS of (14), the number of $-1$s is $0 \leq n \leq m$. $n = 0$ (i.e., all elements in $m$ hence in the original series being $+1$) corresponds to the maximum energy level given by

$$E_{max} = \frac{N(N-1)(2N-1)}{6}. \tag{15}$$

$$E = E_{max} - 4n. \tag{16}$$

From (16), as $n$ is an integer bounded by $\langle 0, n_{max} \leq m \rangle$, and $E_{max}$ is fixed for a particular sequence length $N$, the number of possible energy levels would be fixed and finite. And the factor –4 in the second term in the RHS of (16) would give the difference of 4 between the contiguous energy levels. This completes the proof. ∎

From (16), we see that we would have finite number of discrete energy levels spaced at an interval of 4 for a binary sequence of length $N$. The minimum energy level (objective of the LABS problem, see (5)) would correspond to the $n_{max}$, which would be derived in the following section.

$$E_{min} = E_{max} - 4n_{max}. \tag{17}$$

### 3.2 Minimum energy level

Let's consider we have a binary sequence of length $N_1$, having $n_1$ number of $-1$ and $p_1$ number of $+1$ where $n_1 + p_1 = N_1$, and we construct a second sequence of length $N_2$ by multiplying all combinations of two elements of the first sequence, then $N_2 = \binom{N_1}{2} = \frac{N_1(N_1 - 1)}{2}$. Each $-1$ in the first sequence would produce $p_1$ number of $-1$s in the second sequence. Therefore, the total number of $-1$s in the second sequence

$$n_2 = n_1 p_1 = n_1(N_1 - n_1). \tag{18}$$

To have maximum number of –1s in the second sequence,

$$\frac{dn_2}{dn_1} = N_1 - 2n_1 = 0,$$
$$\Rightarrow n_1 = \frac{N_1}{2} \tag{19}$$

Therefore, to have $n_{2_{max}}$, we need to follow equation (19). If $N_1$ is even then the optimal condition to have $n_{2_{max}}$ follows (19), whereas if $N_1$ is odd then the optimal condition to have $n_{2_{max}}$ is

$$n_1 = \left\{ \frac{N_1-1}{2}, \frac{N_1+1}{2} \right\}. \tag{20}$$

Using (18-20) we get,

$$n_{2_{max}} = \frac{N_1^2}{4}, \quad \text{if } N_1 \text{ is even}$$
$$= \frac{N_1^2-1}{4}, \quad \text{if } N_1 \text{ is odd} \tag{21}$$

Using (21), we would derive the minimum energy levels for a binary sequence of length $N$. For that, we would utilize the following well-known relation for integers,

$$\sum_{i=1}^{n} i^2 = \frac{n(n+1)(2n+1)}{6}. \tag{22}$$

Theorem 3.2. The minimum energy level for a binary sequence of length $N$ ($N$ is even) is given by $\dfrac{N}{2}$.

*Proof*: Considering the energy equation in (11) and the term $Y$ in the RHS of (11), which comes from the multiplication of the combinations of any two elements at each level of autocorrelation $R_k$, $(k = 1, 2, \ldots, N-2)$, we would like to achieve maximum number of $-1$s in $Y$, $n_{max}$, to apply in (17). It is easy to see that we can achieve $n_{max}$ if only if all the individual levels produce the $-1$s maximally. If $N$ is even, then the successive autocorrelation levels would have $N-1$ (odd), $N-2$ (even), …, $2 (= N - (N-2))$ (even) number of elements. That is, we would have $\dfrac{N-2}{2}$ number of odd and even-length sequences from the successive autocorrelation levels contributing maximally towards the total number of $-1$s in $Y$. Therefore, applying (21),

$$\begin{aligned}
n_{max} &= \frac{1}{4}\left[\begin{array}{l}\{(N-1)^2 - 1 + (N-3)^2 - 1 + \ldots + (N-(N-3))^2 - 1\} + \\ \{(N-2)^2 + (N-4)^2 + \ldots + (N-(N-2))^2\}\end{array}\right] \\
&= \frac{1}{4}\left[\{(N-1)^2 + (N-2)^2 + \ldots + 3^2 + 2^2\} - 1 \cdot \frac{(N-2)}{2}\right] \\
&= \frac{1}{4}\left[\{(N-1)^2 + (N-2)^2 + \ldots + 3^2 + 2^2 + 1^2\} - 1^2 - \frac{(N-2)}{2}\right] \qquad (23)\\
&= \frac{1}{4}\left[\frac{(N-1)N(2N-1)}{6} - \frac{N}{2}\right] \\
&= \frac{N(N-2)(2N+1)}{24}.
\end{aligned}$$

Applying (15), (17) and (23), we get

$$\begin{aligned}E_{min} &= E_{max} - 4n_{max} \\ &= \frac{N(N-1)(2N-1)}{6} - \frac{4N(N-2)(2N+1)}{24} \\ &= \frac{N}{2}.\end{aligned} \qquad (24)$$

This completes the proof. ∎

Theorem 3.3. The minimum energy level for a binary sequence of length $N$ ($N$ is odd) is given by $\frac{N-1}{2}$.

*Proof*: Considering the energy equation in (11) and the term $Y$ in the RHS of (11), which comes from the multiplication of the combinations of any two elements at each level of autocorrelation $R_k$, $(k = 1,2,\ldots,N-2)$, we would like to achieve maximum number of $-1$s in $Y$, $n_{max}$, to apply in (17). As before, we can achieve $n_{max}$ only if all the individual levels produce the $-1$s maximally. If $N$ is odd, then the successive autocorrelation levels would have $N-1$ (even), $N-2$ (odd), …, $2(=N-(N-2))$ (even) number of elements. That is, we would have $\frac{N-1}{2}$ number of even and $\frac{N-3}{2}$ number of odd-length sequences from the successive autocorrelation levels contributing maximally towards the total number of $-1$s in $Y$. Therefore, applying (21),

$$n_{max} = \frac{1}{4}\begin{bmatrix}\{(N-1)^2+(N-3)^2+...+(N-(N-2))^2\}+\\ \{(N-2)^2-1+(N-4)^2-1+...+(N-(N-3))^2-1\}\end{bmatrix}$$

$$= \frac{1}{4}\left[\{(N-1)^2+(N-2)^2+...+3^2+2^2\}-1.\frac{(N-3)}{2}\right]$$

$$= \frac{1}{4}\left[\{(N-1)^2+(N-2)^2+...+3^2+2^2+1^2\}-1^2-\frac{(N-3)}{2}\right] \quad (25)$$

$$= \frac{1}{4}\left[\frac{(N-1)N(2N-1)}{6}-\frac{(N-1)}{2}\right]$$

$$= \frac{(N-1)(N+1)(2N-3)}{24}.$$

Applying (15), (17) and (25), we get

$$E_{min} = E_{max} - 4n_{max}$$
$$= \frac{N(N-1)(2N-1)}{6} - \frac{4(N-1)(N+1)(2N-3)}{24} \quad (26)$$
$$= \frac{N-1}{2}.$$

This completes the proof. ∎

Following theorems 3.2 and 3.3 we would get the minimum energy level possible for any binary sequence of length $N$. The corresponding maximum merit factor possible for any binary sequence of length $N$ can be calculated using (4), (24) and (26) as

$$F_{max} = \frac{N^2}{2.\frac{N}{2}} = N, \quad \text{if } N \text{ is even}$$
$$= \frac{N^2}{2.\frac{(N-1)}{2}} = \frac{N^2}{N-1}, \quad \text{if } N \text{ is odd} \quad (27)$$

These bound values were mentioned by Kristiansen in his thesis [19–20]. However, that discussion did not consider the role and contribution of the individual autocorrelation levels towards the minimum energy level as discussed in the derivation of the abovementioned theorems. As we shall see later that the individual autocorrelation level analysis would be helpful for analyzing the Barker sequences as well.

## 4. Application results

Golay found the exact solutions for binary sequences with the highest merit factors using exhaustive searches upto length $N = 60$ [2,7,19]. Golay also described the conjectured asymptotic value of $F = 12.32$ [1] for very long sequences. The list of best sequences found so far and their energy levels, merit factors can be referred to in [21]. However, the case for $N = 13$ is interesting because for $N = 13$, the optimal sequence giving the highest merit factor (found by exhaustive search hence the solution is perfect) is given by $x_{13} = \begin{bmatrix} 1 & 1 & 1 & 1 & 1 & -1 & -1 & 1 & 1 & -1 & 1 & -1 & 1 \end{bmatrix}$, the energy is $E_{13} = 6$ and the merit factor is $F_{13} = 14.083$. It is interesting to note that the merit factor for $N = 13$ is higher than the conjectured value. In fact, it is the only value found so far which lies above the conjectured value. Let's analyze the case of $N = 13$ using the number theory-based analysis. The twelve autocorrelation levels for the $x_{13}$ sequence is shown in Table 1. For the autocorrelation levels, we show the number theory-based analysis in Table 2.

Table 1. Autocorrelation levels for the optimal binary sequence of length 13

| Autocorrelation levels ($N=13$) | Sequence |
|---|---|
| $R_1$ | [1  1  1  1  −1  1  −1  1  −1  −1  −1  −1] |
| $R_2$ | [1  1  1  −1  −1  −1  −1  −1  1  1  1] |
| $R_3$ | [1  1  −1  −1  1  −1  1  1  −1  −1] |
| $R_4$ | [1  −1  −1  1  1  1  −1  −1  1] |
| $R_5$ | [−1  −1  1  1  −1  −1  1  1] |
| $R_6$ | [−1  1  1  −1  1  1  −1] |
| $R_7$ | [1  1  −1  1  −1  −1] |
| $R_8$ | [1  −1  1  −1  1] |
| $R_9$ | [−1  1  −1  1] |
| $R_{10}$ | [1  −1  1] |
| $R_{11}$ | [−1  1] |
| $R_{12}$ | [1] |

Table 2. Number theory-based analysis of the autocorrelation levels of the optimal sequence of length 13

| Autocorr. Level ($R_k$) ($N=13$) | Sequence length ($N-k$) | Expected no. of −1 for $n_{max}$ in $Y$ ($N-k$)/2 | Theoretical $n_{max}$ in $Y$ ($N-k$)²/4, ($N-k$) even ($N-k$)²-1/4, ($N-k$) odd | Actual no. of −1 ($n$) | Actual $n_{max}$ in $Y$ $n.(N-k-n)$ | Theoretical − Actual $n_{max}$ |
|---|---|---|---|---|---|---|
| $R_1$ | 12 | 6 | 36 | 6 | 36 | 0 |
| $R_2$ | 11 | 5.5: 5 or 6 | 30 | 5 | 30 | 0 |
| $R_3$ | 10 | 5 | 25 | 5 | 25 | 0 |
| $R_4$ | 9 | 4.5: 4 or 5 | 20 | 4 | 20 | 0 |
| $R_5$ | 8 | 4 | 16 | 4 | 16 | 0 |
| $R_6$ | 7 | 3.5: 3 or 4 | 12 | 3 | 12 | 0 |
| $R_7$ | 6 | 3 | 9 | 3 | 9 | 0 |
| $R_8$ | 5 | 2.5: 2 or 3 | 6 | 2 | 6 | 0 |
| $R_9$ | 4 | 2 | 4 | 2 | 4 | 0 |
| $R_{10}$ | 3 | 1.5: 1 or 2 | 2 | 1 | 2 | 0 |
| $R_{11}$ | 2 | 1 | 1 | 1 | 1 | 0 |
| $R_{12}$ | 1 | 0.5: 0 or 1 | 0 | 0 | 0 | 0 |

In Table 2, the sequence lengths for each autocorrelation levels are shown in column 2, which indicate even (12), odd (11),… length sequences. These sequences would contribute to the number of –1s in the $Y$ sequence as defined in (11). The third column in Table 2 indicates the expected number of –1s in the each sequence at particular autocorrelation level such that the corresponding number of –1s in $Y$ gets maximum, as per (19-20). The numbers like 5.5 in the third column implies 5 or 6 number of –1s are ok. The fourth column shows the theoretical maximum number of –1s ($n_{max}$) in $Y$ as per (21). In comparison, the fifth column shows the actual number of –1s present in the sequences at each autocorrelation levels shown in Table 1. And the sixth column shows the actual values of $n_{max}$ in $Y$ for these actual number of –1s in the autocorrelation levels. The last column is the difference between the theoretical and the actual number of $n_{max}$ in $Y$ (i.e., difference between column 4 and 6 in Table 1). As all the elements in the last column are zero, then clearly this case of $N=13$ falls under the theoretical minimum energy level as discussed in section 3.2. We also see that as all the autocorrelation levels have the optimal number of –1s in order to produce maximum number of –1s in the $Y$ sequence. Therefore, in aggregate, the $Y$ sequence has the theoretical $n_{max}$, hence the energy is at the theoretical minimum level (see (26)) for odd $N$. Therefore, the energy $E_{13} = \frac{N-1}{2} = 6$, and the merit factor (see (27)),

$$F_{13} = \frac{N^2}{N-1} = \frac{13^2}{13-1} = 14.083.$$

Binary sequence of length 13 is not the only case for the theoretical minimum energy level. $N=4$ also satisfies the theoretical minimum energy level condition. For $N=4$, the optimal sequence is given by $x_4 = \begin{bmatrix} 1 & 1 & 1 & -1 \end{bmatrix}$, energy is $E_4 = 2$, and merit factor $F_4 = 4$ [21]. The autocorrelation levels and the number theory-based analysis are shown in Tables 3-4. From the last column of Table 4 we see that there is no difference between the theoretical (column 4) and the actual (column 6) $n_{max}$ in $Y$. Therefore, the energy is at minimum level for sequence with even length (see (25)), $E_4 = \frac{4}{2} = 2$, and merit factor, $F_4 = 4$.

Table 3. Autocorrelation levels for the optimal binary sequence of length 4

| Autocorrelation levels ($N$=4) | Sequence |
|---|---|
| $R_1$ | $\begin{bmatrix} 1 & 1 & -1 \end{bmatrix}$ |
| $R_2$ | $\begin{bmatrix} 1 & -1 \end{bmatrix}$ |
| $R_3$ | $\begin{bmatrix} -1 \end{bmatrix}$ |

Table 4. Number theory-based analysis of the autocorrelation levels of the optimal sequence of length 4

| Autocorr. Level ($R_k$) ($N$=4) | Sequence length ($N$-$k$) | Expected no. of -1 for $n_{max}$ in $Y$ ($N$-$k$)/2 | Theoretical $n_{max}$ in $Y$ ($N$-$k$)$^2$/4, ($N$-$k$) even ($N$-$k$)$^2$-1/4, ($N$-$k$) odd | Actual no. of -1 ($n$) | Actual $n_{max}$ in $Y$ $n.(N$-$k$-$n$) | Theoretical – Actual $n_{max}$ |
|---|---|---|---|---|---|---|
| $R_1$ | 3 | 1.5: 1 or 2 | 2 | 1 | 2 | 0 |
| $R_2$ | 2 | 1 | 1 | 1 | 1 | 0 |
| $R_3$ | 1 | 0.5: 0 or 1 | 0 | 1 | 0 | 0 |

Three other sequences qualify for the theoretical minimum energy levels. These are given below, and could be verified in the list in [21].

$N = 5$, $x_5 = \begin{bmatrix} 1 & 1 & 1 & -1 & 1 \end{bmatrix}$, $E_5 = \dfrac{5-1}{2} = 2$, $F_5 = \dfrac{5^2}{5-1} = 6.25$ [21]. The autocorrelation levels and the number theory-based analysis are shown in Tables 5–6.

Table 5. Autocorrelation levels for the optimal binary sequence of length 5

| Autocorrelation levels ($N=5$) | Sequence |
|---|---|
| $R_1$ | $\begin{bmatrix} 1 & 1 & -1 & -1 \end{bmatrix}$ |
| $R_2$ | $\begin{bmatrix} 1 & -1 & 1 \end{bmatrix}$ |
| $R_3$ | $\begin{bmatrix} -1 & 1 \end{bmatrix}$ |
| $R_4$ | $\begin{bmatrix} 1 \end{bmatrix}$ |

Table 6. Number theory-based analysis of the autocorrelation levels of the optimal sequence of length 5

| Autocorr. Level ($R_k$) ($N=5$) | Sequence length ($N$-$k$) | Expected no. of -1 for $n_{max}$ in $Y$ ($N$-$k$)/2 | Theoretical $n_{max}$ in $Y$ ($N$-$k$)$^2$/4, ($N$-$k$) even ($N$-$k$)$^2$-1/4, ($N$-$k$) odd | Actual no. of -1 ($n$) | Actual $n_{max}$ in $Y$ $n.(N$-$k$-$n$) | Theoretical – Actual $n_{max}$ |
|---|---|---|---|---|---|---|
| $R_1$ | 4 | 2 | 4 | 2 | 4 | 0 |
| $R_2$ | 3 | 1.5: 1 or 2 | 2 | 1 | 2 | 0 |
| $R_3$ | 2 | 1 | 1 | 1 | 1 | 0 |
| $R_4$ | 1 | 0.5: 0 or 1 | 0 | 0 | 0 | 0 |

$N = 7$, $x_7 = \begin{bmatrix} 1 & 1 & 1 & -1 & -1 & 1 & -1 \end{bmatrix}$, $E_7 = \dfrac{7-1}{2} = 3$, $F_7 = \dfrac{7^2}{7-1} = 8.1667$ [21]. The autocorrelation levels and the number theory-based analysis are shown in Tables 7–8.

Table 7. Autocorrelation levels for the optimal binary sequence of length 7

| Autocorrelation levels ($N=5$) | Sequence |
|---|---|
| $R_1$ | $[1\ \ 1\ \ -1\ \ -1]$ |
| $R_2$ | $[1\ \ -1\ \ 1]$ |
| $R_3$ | $[-1\ \ 1]$ |
| $R_4$ | $[1]$ |

Table 8. Number theory-based analysis of the autocorrelation levels of the optimal sequence of length 7

| Autocorr. Level ($R_k$) ($N=5$) | Sequence length ($N-k$) | Expected no. of -1 for $n_{max}$ in Y ($N-k$)/2 | Theoretical $n_{max}$ in Y ($N-k$)²/4, ($N-k$) even ($N-k$)²-1/4, ($N-k$) odd | Actual no. of -1 ($n$) | Actual $n_{max}$ in Y $n.(N-k-n)$ | Theoretical – Actual $n_{max}$ |
|---|---|---|---|---|---|---|
| $R_1$ | 4 | 2 | 4 | 2 | 4 | 0 |
| $R_2$ | 3 | 1.5: 1 or 2 | 2 | 1 | 2 | 0 |
| $R_3$ | 2 | 1 | 1 | 1 | 1 | 0 |
| $R_4$ | 1 | 0.5: 0 or 1 | 0 | 0 | 0 | 0 |

$N = 11$, $x_{11} = [1\ \ 1\ \ 1\ \ -1\ \ -1\ \ -1\ \ 1\ \ -1\ \ -1\ \ 1\ \ -1]$, $E_{11} = \dfrac{11-1}{2} = 5$,

$F_{11} = \dfrac{11^2}{11-1} = 12.1$ [21]. The autocorrelation levels and the number theory-based analysis are shown in Tables 9–10.

Table 9. Autocorrelation levels for the optimal binary sequence of length 11

| Autocorrelation levels ($N$=11) | Sequence |
|---|---|
| $R_1$ | $[1\ \ 1\ \ -1\ \ 1\ \ 1\ \ -1\ \ -1\ \ 1\ \ -1\ \ -1]$ |
| $R_2$ | $[1\ \ -1\ \ -1\ \ 1\ \ -1\ \ 1\ \ -1\ \ -1\ \ 1]$ |
| $R_3$ | $[-1\ \ -1\ \ -1\ \ -1\ \ 1\ \ 1\ \ 1\ \ 1]$ |
| $R_4$ | $[-1\ \ -1\ \ 1\ \ 1\ \ 1\ \ -1\ \ -1]$ |
| $R_5$ | $[-1\ \ 1\ \ -1\ \ 1\ \ -1\ \ 1]$ |
| $R_6$ | $[1\ \ -1\ \ -1\ \ -1\ \ 1]$ |
| $R_7$ | $[-1\ \ -1\ \ 1\ \ 1]$ |
| $R_8$ | $[-1\ \ 1\ \ -1]$ |
| $R_9$ | $[1\ \ -1]$ |
| $R_{10}$ | $[-1]$ |

Table 10. Number theory-based analysis of the autocorrelation levels of the optimal sequence of length 11

| Autocorr. Level ($R_k$) ($N$=11) | Sequence length ($N$-$k$) | Expected no. of -1 for $n_{max}$ in $Y$ ($N$-$k$)/2 | Theoretical $n_{max}$ in $Y$ ($N$-$k$)²/4, ($N$-$k$) even ($N$-$k$)²-1/4, ($N$-$k$) odd | Actual no. of -1 ($n$) | Actual $n_{max}$ in $Y$ $n.(N$-$k$-$n)$ | Theoretical – Actual $n_{max}$ |
|---|---|---|---|---|---|---|
| $R_1$ | 10 | 5 | 25 | 5 | 25 | 0 |
| $R_2$ | 9 | 4.5: 4 or 5 | 20 | 5 | 20 | 0 |
| $R_3$ | 8 | 4 | 16 | 4 | 16 | 0 |
| $R_4$ | 7 | 3.5: 3 or 4 | 12 | 4 | 12 | 0 |
| $R_5$ | 6 | 3 | 9 | 3 | 9 | 0 |
| $R_6$ | 5 | 2.5: 2 or 3 | 6 | 3 | 6 | 0 |
| $R_7$ | 4 | 2 | 4 | 2 | 4 | 0 |
| $R_8$ | 3 | 1.5: 1 or 2 | 2 | 2 | 2 | 0 |
| $R_9$ | 2 | 1 | 1 | 1 | 1 | 0 |
| $R_{10}$ | 1 | 0.5: 0 or 1 | 0 | 1 | 0 | 0 |

Except these five sequences for $N = \{4,5,7,11,13\}$, all other sequences do not seem to attain the theoretical minimum energy levels. This is because, for all those sequences, as per (17), all the autocorrelation levels do not contribute optimally to produce the maximal $n_{max}$ number of –1s in the $Y$ sequence for the energy equation (11). If in a normal case, the number of –1s in the $Y$ sequence be $n$ which has a deviation of $d$ from the theoretical $n_{max}$, such that

$$n = n_{max} - d, \tag{28}$$

then, the energy equation (11) becomes

$$E = E_{max} - 4n = E_{max} - 4n_{max} + 4d = E_{min} + 4d. \tag{29}$$

From (29), it is evident that for any normal sequence, the energy would be given by the theoretical minimum energy level with an added value coming from the four times the deviations from $n_{max}$. We see an example for $N = 10$.

The optimal sequence for $N = 10$ is given as $x_{10} = \begin{bmatrix} 1 & 1 & 1 & 1 & 1 & -1 & -1 & 1 & -1 & 1 \end{bmatrix}$, the energy $E_{10} = 13$, and the merit factor $F_{10} = 3.8462$ [21]. The theoretical $E_{10\,min} = \dfrac{10}{2} = 5$. Hence, this clearly does not qualify for minimum energy level. We examine the autocorrelation levels and the number theory-based analysis in Tables 11–12.

Table 11. Autocorrelation levels for the optimal binary sequence of length 10

| Autocorrelation levels ($N=10$) | Sequence |
|---|---|
| $R_1$ | $[1\ \ 1\ \ 1\ \ 1\ \ -1\ \ 1\ \ -1\ \ -1\ \ -1]$ |
| $R_2$ | $[1\ \ 1\ \ 1\ \ -1\ \ -1\ \ -1\ \ 1\ \ 1]$ |
| $R_3$ | $[1\ \ 1\ \ -1\ \ -1\ \ 1\ \ 1\ \ -1]$ |
| $R_4$ | $[1\ \ -1\ \ -1\ \ 1\ \ -1\ \ -1]$ |
| $R_5$ | $[-1\ \ -1\ \ 1\ \ -1\ \ 1]$ |
| $R_6$ | $[-1\ \ 1\ \ -1\ \ 1]$ |
| $R_7$ | $[1\ \ -1\ \ 1]$ |
| $R_8$ | $[-1\ \ 1]$ |
| $R_9$ | $[1]$ |

Table 12. Number theory-based analysis of the autocorrelation levels of the optimal sequence of length 10

| Autocorr. Level ($R_k$) ($N=10$) | Sequence length ($N-k$) | Expected no. of -1 for $n_{max}$ in $Y$ ($N-k$)/2 | Theoretical $n_{max}$ in $Y$ ($N-k$)²/4, ($N-k$) even ($N-k$)²-1/4, ($N-k$) odd | Actual no. of -1 ($n$) | Actual $n_{max}$ in $Y$ $n.(N-k-n)$ | Theoretical – Actual $n_{max}$ |
|---|---|---|---|---|---|---|
| $R_1$ | 9 | 4.5: 4 or 5 | 20 | 4 | 20 | 0 |
| $R_2$ | 8 | 4 | 16 | 3 | 15 | 1 |
| $R_3$ | 7 | 3.5: 3 or 4 | 12 | 3 | 12 | 0 |
| $R_4$ | 6 | 3 | 9 | 4 | 8 | 1 |
| $R_5$ | 5 | 2.5: 2 or 3 | 6 | 3 | 6 | 0 |
| $R_6$ | 4 | 2 | 4 | 2 | 4 | 0 |
| $R_7$ | 3 | 1.5: 1 or 2 | 2 | 1 | 2 | 0 |
| $R_8$ | 2 | 1 | 1 | 1 | 1 | 0 |
| $R_9$ | 1 | 0.5: 0 or 1 | 0 | 0 | 0 | 0 |

From the last column of Table 12, it is evident that the total deviation $d = 2$. Therefore, according to (29), the energy would be $E_{10} = \dfrac{10}{2} + 4 \times 2 = 13$. This matches with the actual energy value cited above.

## 4.1 Discussion of results

i. There are five sequences of length $N = \{4,5,7,11,13\}$, which attain the theoretical minimum energy level. This also explains the case $N = 13$, which gives the merit factor higher than the conjectured value of 12.32 by Golay [1]. It is interesting to note that these are exactly the Barker sequences or Barker codes [24]. Barker codes of length 11 and 13 are used in direct-sequence spread spectrum and pulse compression radar systems because of their low autocorrelation properties. This is discussed more in section 4.3.

ii. From (29), the original problem of finding binary sequences with the minimum energy level or highest merit factor (see (5)) could be reformulated as to having minimum number of deviations $d$, i.e.

$$\arg\min\ d. \tag{30}$$

iii. From theorem 3.1, it follows that there are finite number of possible energy levels. Equation (29) supports that, and we see as stated in theorem 3.1 that the energy levels are at a constant interval of 4.

iv. The maximum energy level corresponds to the sequence with all +1s.

v. A possible algorithm could be to start from the maximum energy level and then trying to achieve the theoretical maximum number of −1s at each autocorrelation levels, trying to solve the optimization problem stated in (30).

vi. Another interesting point to note is that, the number of –1s in the optimal sequences of length $N$ tend to be $\leq \frac{N}{2}$. Explanation of this could be that the contribution for –1s in the $Y$ sequence maximally comes from the autocorrelation level 1, having $N-1$ number no. elements. From the theory, the expected number of maximum –1s in the autocorrelation level 1 should be $round\left(\frac{N-1}{2}\right)$. As autocorrelation level 1 comes from the multiplication of the $N-1$ contiguous elements of the original sequence, the number of –1s in the original sequence should be $\leq \frac{N}{2}$. This observation could be particularly helpful towards reducing the computational dimension for an exhaustive search from $2^N$ to $\sum_{k=1}^{round(N/2)} \binom{N}{k}$.

For example, for $N = 40$, the exhaustive search space of about $1.0995 \times 10^{12}$ would be reduced to about $6.1868 \times 10^{11}$, a 44% reduction in computation.

## 4.2 Sequences longer than $N = 60$

Sequences upto length 60 [2,3,55] have been found by exhaustive searches, they are the exact solutions for the minimum energy levels. Using (29), we get the exact values of the deviations upto length 60 (see Table 13). Following this, in Fig. 1, we fit a second-order polynomial curve upto the length 60 in plot (i), and extrapolate the curve upto length 304 and compare against the deviations of the best found sequences in plot (ii).

Table 13. Exact deviation values upto sequence length 60

| Sequence length | Best found energy | Theoretical energy | Deviation |
|---|---|---|---|
| 4 | 2 | 2 | 0 |
| 5 | 2 | 2 | 0 |
| 6 | 7 | 3 | 1 |
| 7 | 3 | 3 | 0 |
| 8 | 8 | 4 | 1 |
| 9 | 12 | 4 | 2 |
| 10 | 13 | 5 | 2 |
| 11 | 5 | 5 | 0 |
| 12 | 10 | 6 | 1 |
| 13 | 6 | 6 | 0 |
| 14 | 19 | 7 | 3 |
| 15 | 15 | 7 | 2 |
| 16 | 24 | 8 | 4 |
| 17 | 32 | 8 | 6 |
| 18 | 25 | 9 | 4 |
| 19 | 29 | 9 | 5 |
| 20 | 26 | 10 | 4 |
| 21 | 26 | 10 | 4 |
| 22 | 39 | 11 | 7 |
| 23 | 47 | 11 | 9 |
| 24 | 36 | 12 | 6 |
| 25 | 36 | 12 | 6 |
| 26 | 45 | 13 | 8 |
| 27 | 37 | 13 | 6 |
| 28 | 50 | 14 | 9 |
| 29 | 62 | 14 | 12 |
| 30 | 59 | 15 | 11 |
| 31 | 67 | 15 | 13 |
| 32 | 64 | 16 | 12 |
| 33 | 64 | 16 | 12 |
| 34 | 65 | 17 | 12 |
| 35 | 73 | 17 | 14 |
| 36 | 82 | 18 | 16 |
| 37 | 86 | 18 | 17 |
| 38 | 87 | 19 | 17 |
| 39 | 99 | 19 | 20 |
| 40 | 108 | 20 | 22 |
| 41 | 108 | 20 | 22 |
| 42 | 101 | 21 | 20 |
| 43 | 109 | 21 | 22 |
| 44 | 122 | 22 | 25 |
| 45 | 118 | 22 | 24 |
| 46 | 131 | 23 | 27 |
| 47 | 135 | 23 | 28 |
| 48 | 140 | 24 | 29 |
| 49 | 136 | 24 | 28 |
| 50 | 153 | 25 | 32 |
| 51 | 153 | 25 | 32 |
| 52 | 166 | 26 | 35 |
| 53 | 170 | 26 | 36 |
| 54 | 175 | 27 | 37 |
| 55 | 171 | 27 | 36 |
| 56 | 192 | 28 | 41 |
| 57 | 188 | 28 | 40 |
| 58 | 197 | 29 | 42 |
| 59 | 205 | 29 | 44 |
| 60 | 218 | 30 | 47 |

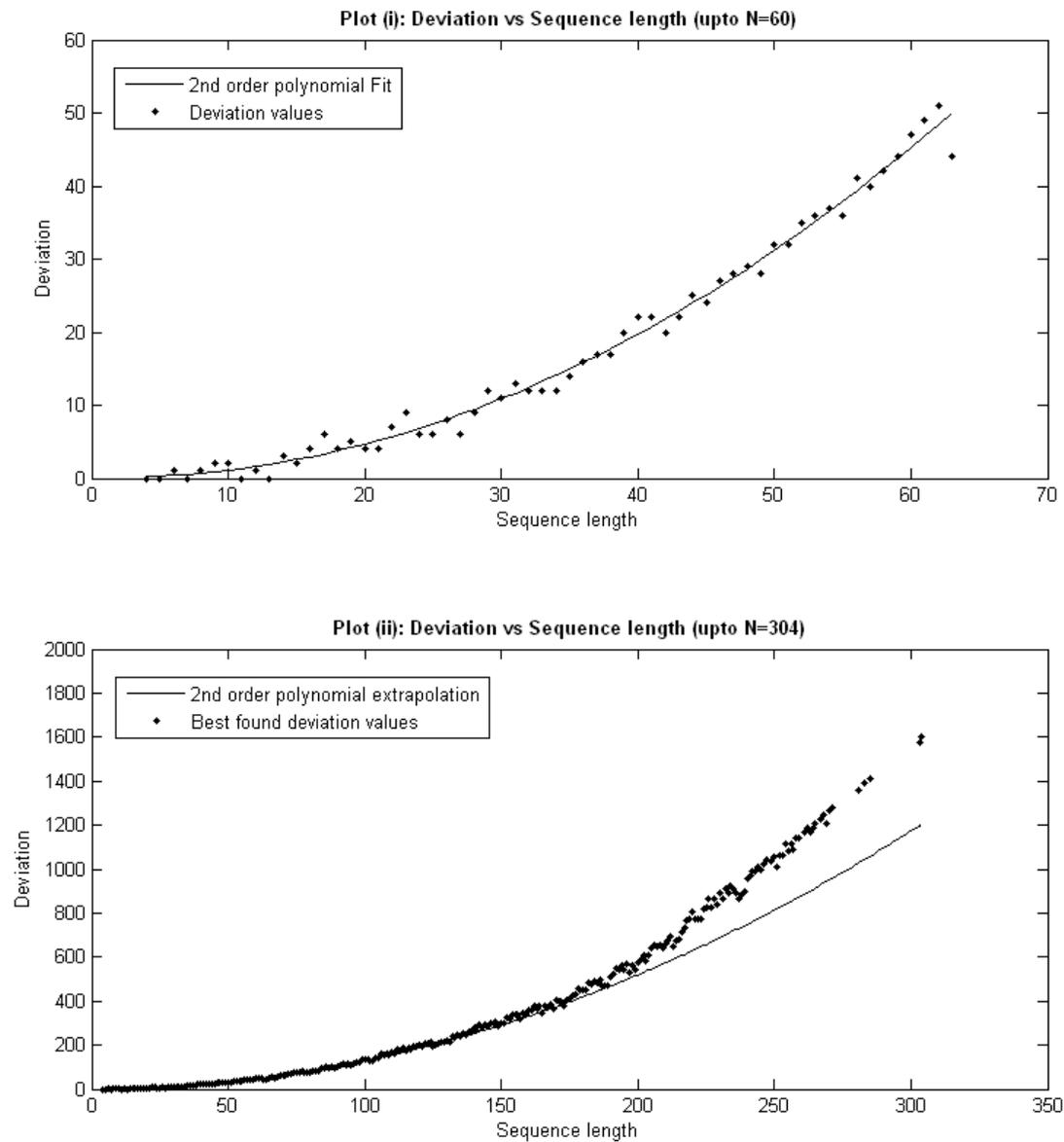

Fig. 1. Second-order polynomial fit for deviation against sequence length upto 60 (plot i) and extrapolated upto length 304 in comparison with best found solutions (plot ii).

From Fig. 1, plot (ii), we see that the extrapolated polynomial curve tend to deviate from the best found deviation values [21] from around sequence length $N = 190$. Therefore, we can empirically conclude that from the sequence length of 190 onwards the best found sequences as listed in [21] are not optimal, hence better sequences with lower number of deviations hence lower energy levels might exist.

Some experimental results for sequences longer than $N = 60$ are discussed by Halim et al [22]. A visualization tool, *Viz*, for the stochastic search of longer sequences can be found at [23].

### 4.3 Comments on Barker sequences

From the examples in section 4, it is evident that five sequences of length $N = \{4,5,7,11,13\}$ attain the theoretical minimum energy level. Interestingly, these are exactly the Barker sequences or Barker codes [24]. It is conjectured that Barker sequences exist only when $N$ is prime and $N \leq 13$. Storer and Turyn [25] proved this conjecture for all odd $N$.

From the analysis of the minimum energy level in section 3.2, we see that a binary sequence can only attain the minimum energy level if and only if all the autocorrelation levels attain their individual lowest energy level. This is a very special condition, and only five sequences obey this (the trivial cases of $N = 2,3$ also satisfy this). In reality, the energy of the sequences would follow (29), with a deviation parameter $d$. Therefore, to have minimum energy level, or Barker sequences, we must satisfy the condition

$$d = 0.\qquad(31)$$

However, from Table 13 and Fig. 1, we can notice that the deviation parameter is almost a monotonically increasing second-order polynomic function of the sequence length. Therefore, to have Barker sequences for $N > 13$, the curves shown in Fig. 1 must jump sharply to zero to satisfy (31). Physically, this seems rather unrealistic, substantiating the conjecture.

Furthermore, for a binary sequence of length $N$ and merit factor $F$, from (4 & 29) we have,

$$d = \frac{1}{4}\left[\frac{N^2}{2F} - E_{\min}\right]. \tag{32}$$

Using (24 & 26),

$$d = \frac{N^2 - FN}{8F} \quad \text{for even } N$$
$$= \frac{N^2 - FN + F}{8F} \quad \text{for odd } N \tag{33}$$

Applying the condition in (31), to produce minimum energy level or Barker sequences, we have the following conditions, considering the fact that the merit factor $F$ cannot physically attain infinitely large value.

$$N^2 - FN = 0 \implies N = F \quad \text{for even } N, \tag{34}$$

$$N^2 - FN + F = 0 \Rightarrow N = \frac{F \pm \sqrt{F^2 - 4F}}{2} \qquad for\ odd\ N. \tag{35}$$

Considering Golay's conjectured asymptotic value of $F = 12.32$, for the Barker sequences, from (34) we get the value of $N = 12.32$, and from (35) $N = 11.2222, 1.0978$. This strongly supports the conjecture on the existence of the Barker sequences only for $N \leq 13$. Nevertheless, the analysis in this paper shows that the Barker sequences are the best possible low autocorrelation binary sequences, achieving the theoretical minimum energy level.

## 5. Conclusions

Low autocorrelation binary sequences are very important for communication applications. And it is a notoriously difficult computational problem to find binary sequences with low aperiodic autocorrelations. The problem can also be stated in terms of finding binary sequences with minimum energy levels or maximum merit factor defined by MJE Golay [1]. Golay also described the conjectured asymptotic value of the merit factor $F = 12.32$ [1] for very long sequences.

In this paper, a theorem has been proven to show that the possible energy levels for the binary sequence of a particular length are of finite numbers and spaced at an equal interval of 4. Two more theorems are proven to show that the theoretical minimum energy level of a binary sequence of even and odd length of $N$ are $\frac{N}{2}$ and $\frac{N-1}{2}$ respectively. Application results using the number theory-based analysis show that there are five sequences of length $N = \{4,5,7,11,13\}$, which attain the theoretical minimum energy level. This also successfully explains the case of $N = 13$, for which the merit factor is 14.083 higher than the conjectured value. It is interesting to note that these are exactly the Barker sequences [24] used for pulse compression of radar signals.

Analysis of other sequences show that the minimum energy level possible is given by the theoretical minimum energy level plus four times a deviation parameter, indicating deviation from the theoretical minimum energy level. Hence, the problem could be restated to minimization of the deviation parameter. Knowing the minimum energy level reveals the optimal number of –1s to be present in the sequence, it could be particularly useful for reducing the search space for an exhaustive search.

Analysis of the deviation parameter for sequences upto length 60, which have been found to be the exact solutions through exhaustive searches, empirically shows that the best found sequences [21] are not optimal beyond the length of 190. Hence, empirically better sequences with lower energy levels might exist beyond the length of 190. Furthermore, the approximately monotonically increasing polynomic curve of the deviation parameter versus sequence length (>60) strongly supports the conjecture that Barker sequences only exists for sequences of length 13 or less. This is because, to have Barker sequences for sequence length greater than 13, the deviation parameter curve has to sharply come down to zero, which is physically rather unrealistic.

**Acknowledgments**

The author would like to thank Dr. Jakob Bernasconi for numerous encouraging discussions, and the anonymous reviewers for the constructive review, helping to upgrade the paper.

# Appendix

*Proof*: $\sum_{k=1}^{N-2} \binom{N-k}{2} = \frac{N(N-1)(N-2)}{6}$.

We know that for integers, $\sum_{i=1}^{n} i = \frac{n(n+1)}{2}$, $\sum_{i=1}^{n} i^2 = \frac{n(n+1)(2n+1)}{6}$.

$$\sum_{k=1}^{N-2} \binom{N-k}{2} = \frac{1}{2}[(N-1)(N-2) + (N-2)(N-3) + \ldots + 2.1]$$
$$= \frac{1}{2}[(N-1)(N-1-1) + (N-2)(N-2-1) + \ldots + 2.(2-1)]$$
$$= \frac{1}{2}[\{(N-1)^2 + (N-2)^2 + \ldots + 2^2\} - \{(N-1) + (N-2) + \ldots + 2\}]$$
$$= \frac{1}{2}[\{(N-1)^2 + (N-2)^2 + \ldots + 2^2 + 1^2\} - \{(N-1) + (N-2) + \ldots + 2 + 1\}]$$
$$= \frac{1}{2}\left[\frac{(N-1)N(2N-1)}{6} - \frac{(N-1)N}{2}\right]$$
$$= \frac{N(N-1)(N-2)}{6}.$$

# Biography

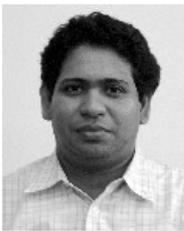

**Abhisek Ukil** received the B.E. degree in electrical engineering from the Jadavpur Univ., Calcutta, India, in 2000 and the M.Sc. degree in electronic systems and engineering management from the Univ. of Bolton, Bolton, UK, and Southwestphalia Univ. of Applied Sciences, Soest, Germany in 2004. He was a DAAD scholar during M.Sc. He received his Ph.D. from the Tshwane Univ. of Technology, Pretoria, South Africa in 2006.

Since 2006, he is a research scientist at the 'Integrated Sensor Systems group,' ABB Corporate Research Center in Baden-Daettwil, Switzerland. He has published over 30 scientific papers, including a monograph *Intelligent Systems and Signal Processing in Power Engineering* by Springer, Heidelberg in 2007. His research interests include signal processing, sequences, machine learning, power systems, embedded systems, number theory.